\documentclass{ieeeaccess}
\usepackage{cite}
\usepackage{amsmath,amssymb,amsfonts}
\usepackage{algorithmic}
\usepackage{graphicx}
\usepackage{textcomp}
\usepackage{csquotes}
\usepackage{soul}
\usepackage{subcaption}

\usepackage[inline,nomargin,marginclue,status=draft]{fixme}
\fxsetup{envlayout=color,theme=color}
\FXRegisterAuthor{mf}{adb}{MF}
\FXRegisterAuthor{mi}{ami}{MI}
\FXRegisterAuthor{ap}{aap}{AP}

\def\BibTeX{{\rm B\kern-.05em{\sc i\kern-.025em b}\kern-.08em
    T\kern-.1667em\lower.7ex\hbox{E}\kern-.125emX}}
\begin{document}
\history{Date of publication xxxx 00, 0000, date of current version xxxx 00, 0000.}
\doi{}

\newcommand{\prnu}{PRNU }
\newcommand{\pce}{\textit{PCE}}
\newcommand{\fpr}{FPR }
\newcommand{\tpr}{TPR }
\newcommand{\nua}{NUA}
\newcommand{\nuas}{NUAs}

\title{A leak in PRNU based source identification.
Questioning fingerprint uniqueness}
\author{\uppercase{Massimo Iuliani}\authorrefmark{1,2},
\uppercase{Marco Fontani\authorrefmark{3}, and Alessandro Piva}\authorrefmark{1,2},
\IEEEmembership{Fellow, IEEE}}
\address[1]{Department of Information Engineering, University of Florence, Via di S. Marta, 3, 50139 Florence, Italy}
\address[2]{FORLAB, Multimedia Forensics Laboratory, PIN Scrl, Piazza G. Ciardi, 25, 59100 Prato, Italy}
\address[3]{AMPED Software, Loc. Padriciano 99, 34149 Trieste, Italy}

\markboth
{Massimo Iuliani \headeretal: A leak in PRNU based source identification. Questioning fingerprint uniqueness}
{Massimo Iuliani \headeretal: A leak in PRNU based source identification. Questioning fingerprint uniqueness}

\corresp{Corresponding author: Alessandro Piva (e-mail: alessandro.piva@unifi.it).}

\begin{abstract}
Photo Response Non-Uniformity (PRNU) is considered the most effective trace for the image source attribution task. Its uniqueness ensures that the sensor pattern noises extracted from different cameras are strongly uncorrelated, even when they belong to the same camera model. 
However, with the advent of computational photography, most recent devices heavily process the acquired pixels, possibly introducing non-unique artifacts that may reduce PRNU noise's distinctiveness, especially when several exemplars of the same device model are involved in the analysis. Considering that PRNU is an image forensic technology that finds actual and wide use by law enforcement agencies worldwide, it is essential to keep validating such technology on recent devices as they appear.
In this paper, we perform an extensive testing campaign on over 33.000 Flickr images belonging to $45$ smartphone and $25$ DSLR camera models released recently to determine how widespread the issue is and which is the plausible cause. 
Experiments highlight that most brands, like Samsung, Huawei, Canon, Nikon, Fujifilm, Sigma, and Leica, are strongly affected by this issue.
We show that the primary cause of high false alarm rates cannot be directly related to specific camera models, firmware, nor image contents.
It is evident that the effectiveness of \prnu based source identification on the most recent devices must be reconsidered in light of these results. 
Therefore, this paper is intended as a call to action for the scientific community rather than a complete treatment of the subject.
Moreover, we believe publishing these data is important to raise awareness about a possible issue with PRNU reliability in the law enforcement world.
\end{abstract}

\titlepgskip=-15pt

\maketitle

\section{Introduction}
\label{sec:intro}
Photo Response Non-Uniformity (\prnu) is considered the most distinctive trace to link an image to its originating device~\cite{lukas2006digital}. 
Such a trace has been studied and improved for more than a decade and can be used for several tasks: (i) attribute an image to its source camera~\cite{chen2008determining}; (ii) determine whether two images belong to the same camera~\cite{goljan2007identifying}; (iii) cluster a large number of images based on the originating device~\cite{marra2017blind}; (iv) determine whether a photo and a video have been captured with the same camera~\cite{iuliani2019hybrid}; (v) detect and localize the presence of a manipulated image region~\cite{chen2008determining}.

After its introduction~\cite{lukas2006digital}, several refinements were proposed to improve the usage of the \prnu trace under challenging scenarios. 
Non-unique artifacts introduced by color interpolation and JPEG compression were studied and removed~\cite{chen2008determining}.
a more general approach was proposed to manage the case of cropped and resized images, and the peak-to-correlation-energy (PCE) ratio was introduced as a more robust way to measure the peak value of the correlation ~\cite{goljan2008camera},~\cite{goljan2009large}. To further improve its effectiveness and efficiency, several filters and pre-processing steps have been designed ~\cite{amerini2009analysis},~\cite{li2010source},~\cite{lin2015preprocessing},~\cite{al2016spn},~\cite{lawgaly2016sensor}, and, recently, also data-driven approaches have been proposed \cite{kirchner2019spn,mandelli2020cnn}. \prnu compression techniques have been developed to enable very large scales operations~\cite{valsesia2015large},~\cite{bondi2018improving}, previously impossible due to the size of the \prnu pattern and the matching operations' complexity. 
Such a trace has also been studied under more complicated setups, e.g., when the media is exchanged through social media~\cite{shullani2017vision,caldelli2018prnu}, or when it is acquired using digital zoom~\cite{goljan2008camera}. \prnu has also been combined with camera-model fingerprints to increase the performance of source identification \cite{cozzolino2020combining}.

All the mentioned works have been carried out under scenarios where the images used in the experiments were taken by camera devices or smartphones that follow the standard acquisition pipeline~\cite{piva2013overview}. 
The first step towards more modern acquisition devices was the study of \prnu detection in the presence of Electronic Image Stabilization (EIS).
EIS introduces a pixel grid misalignment that, if not taken into account, leads \prnu based testing to failure~\cite{taspinar2016source,mandelli2019facing, mandelli2020modified,bellavia2021experiencing}.
Anyhow, under all the above scenarios, the main risk was mainly related to the increase of the false negative rate.
On the other hand, when dealing with false positive rate, the effectiveness of \prnu has never been put in doubt.
Indeed, the state of the art~\cite{shullani2017vision} highlights that, when a good reference is available, the source identification task on a single image can achieve an area under curve (AUC) of $0.99$ and a false positive rate below $0.001$ (at the commonly agreed PCE ratio threshold of $60$).

Thanks to these results and similar experiments carried out with other datasets, forensic experts, law enforcement agencies, and the research community agree on the effectiveness of PRNU as a means for the source identification task. 
More specifically, a false positive is expected to be a sporadic event, thus assuring that a high PCE ratio value represents a robust and reliable finding.

This belief, actually, is mainly based on results achieved on devices released several years ago (such as those in the VISION dataset~\cite{shullani2017vision}), so their imaging technology is somewhat obsolete.  

With the advent of computational photography, new challenges appear since the image acquisition pipeline is rapidly changing, and it appears to be strongly customized by each brand. Novelties involve both the acquisition process, e.g., with the use of pixel binning~\cite{bock2006apparatus, agranov2017pixel}, and the in-camera processing, with the development of customized HDR algorithms~\cite{Eilertsen2017HDR} and the exploitation of artificial intelligence for a variety of image improvements \cite{Chen_2018_CVPR,zhang2019zoom}.

These new customized pipelines can introduce new non-unique artifacts (\nua), shared among different cameras of the same model, or even among cameras of different models, disturbing the \prnu extraction process.
This fact poses the severe risk that images belonging to different cameras expose correlated patterns, thus increasing the false alarm rate abruptly in \prnu detection. 
This is a severe issue since \prnu based source identification is currently used as evidence in court\footnote{(United States of America v Nathan Allen Railey, United States District Court: Southern District of Alabama: 1:10-cr-00266-CG-N, 2011)} and is implemented in several forensic software supplied to law enforcement agencies and intelligence services, like \prnu Compare Professional \footnote{https://www.forensicinstitute.nl/products-and-services/forensic-products} developed by the Netherlands Forensic Institute (NFI), and Amped Authenticate\footnote{https://ampedsoftware.com/authenticate} by Amped Software.

To the best of our knowledge, this paper represents the first study where the \prnu based source identification is tested under a large set of images taken by the most recent devices that exploit the newest imaging technologies.
In particular, we highlight that this forensic technique, applied as it is on modern cameras, is not reliable anymore since it is strongly affected by unexpected high correlations among different devices of the same camera model or brand.
Considering that \prnu based source camera identification is currently used by law enforcement agencies worldwide, often to investigate serious crimes such as child sexual exploitation, we believe it is fundamental that the scientific community cross-verifies the results we obtained (the dataset presented in the paper is made available) and tries to shed light on this potentially disruptive discovery as promptly as possible. 

The paper is organized as follows: Section~\ref{sec:theory} summarizes the theoretical framework and the main pipeline for the \prnu extraction and comparison; in Section~\ref{sec:data} presents the collected dataset; in Section~\ref{sec:experiments} we describe the experiments performed on the acquired data and comment on the results; in Section~\ref{sec:discussion} we focus on the results obtained comparing different exemplars of two specific devices. Section~\ref{sec:conclusions} draws some conclusions and highlights questions that remain open.

Everywhere in this paper, vectors and matrices are indicated in bold as $\mathbf{X}$ and their components as $\mathbf{X}(i)$ and $\mathbf{X}(i,j)$ respectively. All operations are element-wise unless mentioned otherwise. 
Given two vectors $\mathbf{X}$ and $\mathbf{Y}$, $||\mathbf{X}||$ is the euclidean norm of $\mathbf{X}$, $\mathbf{X \cdot Y}$ is the dot product between $\mathbf{X}$ and $\mathbf{Y}$, $\bar{\mathbf{X}}$ is the mean value of $\mathbf{X}$, $\rho(s_1,s_2; \mathbf{X},\mathbf{Y})$ is the normalized cross-correlation between $\mathbf{X}$ and $\mathbf{Y}$ calculated at the spatial shift $(s_1,s_2)$ as

\small{
\begin{equation*}
\rho(s_1,s_2; \mathbf{X},\mathbf{Y}) = \frac{\sum_{i} \sum_{j} ( \mathbf{X}[i,j] - \bar{\mathbf{X}} ) ({\mathbf{Y}[i+s_1,j+s_2] - \bar{\mathbf{Y}}} )} {||\mathbf{X} - \bar{\mathbf{X}}|| ||\mathbf{Y} - \bar{\mathbf{Y}}||},
\end{equation*}
}

\normalsize

where the shifts $[i,j]$ and $[i+s_1,j+s_2]$ are taken modulo the horizontal and vertical image dimensions\footnote{If $\mathbf{X}$ and $\mathbf{Y}$ dimensions mismatch, a zero down-right padding is applied.}.
Furthermore, we denote maximum by $\rho(\mathbf{s}_{peak}; \mathbf{X},\mathbf{Y}) = \max\limits_{{s_1,s_2}} \rho(s_1,s_2; \mathbf{X},\mathbf{Y})$.
The notations are simplified in $\rho(\mathbf{s})$ and in $\rho(\mathbf{s}_{peak})$ when the two vectors cannot be misinterpreted.

%
\section{\prnu based source identification}
\label{sec:theory}

\prnu defines a subtle variation among pixels amplitude due to the different sensitivity to light of the sensor's elements. 
This defect introduces a unique fingerprint into every image the camera takes.
Then, camera fingerprints can be estimated and compared against images to determine the originating device.
A camera fingerprint is usually estimated from $n$ images $\mathbf{I}_1, \dots, \mathbf{I}_n$ as follows:
a denoising filter~\cite{lukas2006digital},~\cite{mihcak1999spatially} is applied to the images to obtain the noise residuals $\mathbf{W}_1, \dots, \mathbf{W}_n$.
The camera fingerprint estimate $\widetilde{\mathbf{K}}$ is derived by the maximum likelihood
estimator~\cite{chen2008determining}:
\begin{equation}
\widetilde{\mathbf{K}} = \frac{\sum_{i=1}^N \mathbf{W}_{i} \mathbf{I}_{i}}{\sum_{i=1}^N \mathbf{I}_{i}^2}.
\label{eq:lr}
\end{equation}
Two further processing are applied to $\widetilde{\mathbf{K}}$ to remove demosaicing traces, JPEG blocking and other non-unique artifacts~\cite{chen2008determining}.

The most common source identification test tries to determine whether a query image $\mathbf{I}$ belongs to a specific camera. 
Given $\mathbf{W}$ the noise residual extracted from $\mathbf{I}$ and the reference camera fingerprint estimate $\widetilde{\mathbf{K}}$, 
the two-dimensional normalized cross-correlation $\rho(s_1,s_2; \mathbf{X}, \mathbf{Y})$ is computed with 
$\mathbf{X}=\mathbf{I} \widetilde{\mathbf{K}}, \mathbf{Y}=\mathbf{W}$ for any plausible shift $(s_1,s_2)$; then the peak-to-correlation energy (PCE) ratio~\cite{goljan2009large} is derived as
\begin{equation}
\label{eq:PCE}
PCE = \frac{\rho(\mathbf{s}_{peak})^2} { \frac{1}{mn - |\mathcal{V}|} \sum\limits_{\mathbf{s}\notin\mathcal{V}} \rho(\mathbf{s})^2 }
\end{equation}
where $\mathcal{V}$ is a small set of peak neighbours and $(m,n)$ is the image pixel resolution.
When $PCE>\tau$, for a given threshold $\tau$, we decide that $\mathbf{W}$ is found within $\mathbf{I}$, i.e. the image belongs to the reference camera. 
A threshold of $60$ is commonly accepted by the research community since, according to experimental validation, it guarantees a negligible false alarm rate (FAR)~\cite{goljan2009large},~\cite{taspinar2016source},~\cite{entrieri2016patch},~\cite{mandelli2019facing}.

\section{Collected Dataset}
\label{sec:data}

In order to understand the impact of the technological novelties in the imaging field on \prnu detection, we collected images from Flickr. Indeed, this image sharing platform allows, under some settings, to download the original version of shared images \cite{giudice2017classification}.
We downloaded images belonging to $45$ smartphone and $25$ DSLR camera models chosen among the most widespread in the market\footnote{based on statistics found on https://www.statista.com/}.
The lists of smartphones and cameras are provided in Table~\ref{tb:smartphones} and Table~\ref{tb:cameras} respectively, together with the number of exemplars for each model.
For each targeted device model, we aimed to download images from 10 different users, ensuring that at least 70 images were available for each user.
For some devices, especially very recent ones, we could only find sufficient data for a lower number of users.

\begin{table}[h!]
\caption{List of smartphones composing the Flickr dataset. For each model, the number of Flickr users is shown. The devices for which multiple resolutions were considered are highlighted in bold.}
\centering 
\begin{tabular}{ llll }
\textbf{ID} & \textbf{Brand} & \textbf{Model} & \textbf{Exemplars} \\ \hline
	S01 & Apple & iPhone 11 & 8 \\ 
	S02 & Apple & iPhone 11 Pro & 7 \\ 
	S03 & Apple & iPhone 11 Pro Max & 9 \\ 
	S04 & Apple & iPhone 6 & 10 \\ 
	S05 & Apple & iPhone 6 Plus & 10 \\ 
	S06 & Apple & iPhone 7 & 10 \\ 
	S07 & Apple & iPhone 7 Plus & 7 \\ 
	S08 & Apple & iPhone 8 & 10 \\ 
	S09 & Apple & iPhone 8 plus & 10 \\ 
	S10 & Apple & iPhone X & 8 \\ 
	S11 & Apple & iPhone XR & 5 \\ 
	S12 & Apple & iPhone XS & 7 \\ 
	S13 & Apple & iPhone XS Max & 9 \\ 
	S14 & Huawei & P20 Lite & 10 \\ 
	S15 & Huawei & P20 Pro & 2 \\ 
	\textbf{S16} & \textbf{Huawei} & \textbf{P30} & 11 \\ 
	\textbf{S17} & \textbf{Huawei} & \textbf{Mate 20 Pro} & 19 \\ 
	\textbf{S18} & \textbf{Huawei} & \textbf{P30 Lite}  & 11 \\ 
	S19 & Huawei & P Smart 2019 & 10 \\ 
	S20 & Huawei & Mate 20 Lite & 10 \\ 
	\textbf{S21} & \textbf{Huawei} & \textbf{P30 Pro}  & 12\\ 
	\textbf{S22} & \textbf{Huawei} & \textbf{P10 } & 19\\ 
	S23 & Motorola & E5 Play & 9 \\ 
	S24 & Nokia & PureView 808 & 8 \\ 
	S25 & Oneplus & 6 & 10 \\ 
	S26 & Oneplus & 6T & 8 \\ 
	S27 & Oppo & A9 2020 & 3 \\ 
	S28 & Realme & C2 & 2 \\ 
	S29 & Samsung & Galaxy A10e & 4 \\ 
	S30 & Samsung & Galaxy A20 & 3 \\ 
	S31 & Samsung & Galaxy A40 & 5 \\ 
	S32 & Samsung & Galaxy A50 & 3 \\ 
	S33 & Samsung & Galaxy S6 & 8 \\ 
	S34 & Samsung & Galaxy S7 & 8 \\ 
	S35 & Samsung & Galaxy S7 edge & 9 \\ 
	S36 & Samsung & Galaxy S8 & 8 \\ 
	S37 & Samsung & Galaxy S8+ & 10 \\ 
	S38 & Samsung & Galaxy S9 & 7 \\ 
	S39 & Samsung & Galaxy S9+ & 9 \\ 
	S40 & Samsung & Galaxy S10e & 5 \\
	\textbf{S41} & \textbf{Samsung} & \textbf{Galaxy S10} & 10 \\ 
	\textbf{S42} & \textbf{Samsung} & \textbf{Galaxy S10+} & 10 \\ 
	S43 & Xiaomi & Mi 9 & 7 \\ 
	S44 & Xiaomi & Mi A3 & 3 \\ 
	\textbf{S45} & \textbf{Xiaomi} & \textbf{Redmi Note 7} & 9 \\ 
\end{tabular}
\label{tb:smartphones}
\end{table}

\begin{table}[h!]
\caption{List of cameras composing the Flickr dataset. For each model, the number of Flickr users is shown.}
\centering 
\begin{tabular}{ llll }
	\textbf{ID} & \textbf{Brand} & \textbf{Model} & \textbf{Exemplars} \\ \hline
	C01 & Canon & EOS 90D & 6 \\ 
	C02 & Canon & EOS M6 Mark II & 5 \\ 
	C03 & Canon & EOS Rebel SL3 & 7 \\ 
	C04 & Canon & EOS RP & 6 \\ 
	C05 & Canon & EOS-1D X Mark III & 2 \\ 
	C06 & Canon & PowerShot G7 X Mark II& 9 \\ 
	C07 & Canon & PowerShot G7 X Mark III & 2 \\ 
	C08 & Nikon  & Coolpix A1000 & 2 \\ 
	C09 & Sony & DSC-Rx0 & 4 \\ 
	C10 & Sony & DSC-Rx100m7 & 4 \\ 
	C11 & Olympus  & E-M1X & 6 \\ 
	C12 & Gopro & Hero8 black & 2 \\ 
	C13 & Sony & ILCE-6400 & 7 \\ 
	C14 & Sony & ILCE-6600 & 2 \\ 
	C15 & Sony & ILCE-7rm4 & 3 \\ 
	C16 & Sony & ILCE-7s & 8 \\ 
	C17 & Sony & ILCE-9 & 5 \\ 
	C18 & Leica  & Q2 & 6 \\ 
	C19 & Nikon  &  D780 & 2 \\ 
	C20 & Nikon  &  Z50 & 3 \\ 
	C21 & Sigma & fp & 4 \\ 
	C22 & Olympus  & TG-6 & 8 \\ 
	C23 & Leica & V-Lux (typ 114) & 4 \\ 
	C24 & Fujifilm & X-Pro3 & 2 \\ 
	C25 & Fujifilm & X-T30 & 5 \\ 
\end{tabular}
\label{tb:cameras}
\end{table}

These rules regulated the download and further selection of images:
\begin{enumerate}
\item Exif \textit{Make} and \textit{Model} metadata must be present and match the targeted device;
\item image resolution must match the maximum resolution allowed by the device.
For some devices, especially those featuring pixel binning, we downloaded multiple resolutions.
In such cases, ten users were targeted for each resolution;
\item image metadata must not contain traces of processing software. To achieve this, we checked the \textit{Exif software} tag against a blacklist of over $90$ image processing software;
\end{enumerate}
We then analyzed images for each user and computed the mode of their \textit{Focal Length} metadata, when available. We maintained only images whose focal length matched the mode, while others were discarded.
This further filtering rule ensures that images from the same camera are used in the case of multi-camera devices. Finally, we divided the remaining images into two groups:
\begin{itemize}
    \item \textit{Reference images}: a group of no less than 20 and no more than 35 images, for which the \textit{Digital Zoom} metadata was either absent or equal to 1.0;
    \item \textit{Test images}: all remaining images for the user.
\end{itemize}
Overall, the dataset comprises 24.908 images from $372$ smartphone and 8.347 images from $114$ DSLR camera exemplars.
Table \ref{tab:summary} reports the main statistics of the dataset.
The collected dataset will be made available to researchers upon request to the corresponding author.

\section{Experiments}
\label{sec:experiments}

For each collected device, we carried out the source attribution task according to the following protocol:
\begin{itemize}
\item \textbf{Reference building}: a camera fingerprint was created for each user, using the available \textit{Reference images};
\item \textbf{Match test}: all available $N_t$ \textit{Test images} from the same user were compared against the fingerprint, testing all possible rigid rotations ($0^{\circ}$, $90^{\circ}$, $180^{\circ}$, $270^{\circ}$);  
\item \textbf{Mismatch test}: a set of $200-N_t$ images captured with the same device model but by a different user were compared against the fingerprint, testing all possible rigid rotations and scaling them to match the fingerprint size if needed;
\end{itemize}

\begin{figure*}
\centering
\caption{PCE statistics computed from Apple, Huawei, and Samsung devices (top, middle and bottom plot respectively).
Matching and mismatching images are reported in green and red, respectively. The threshold of $60$ is highlighted by the red dashed line.}
  \begin{tabular}{cc}
    \includegraphics[width=.9\textwidth]{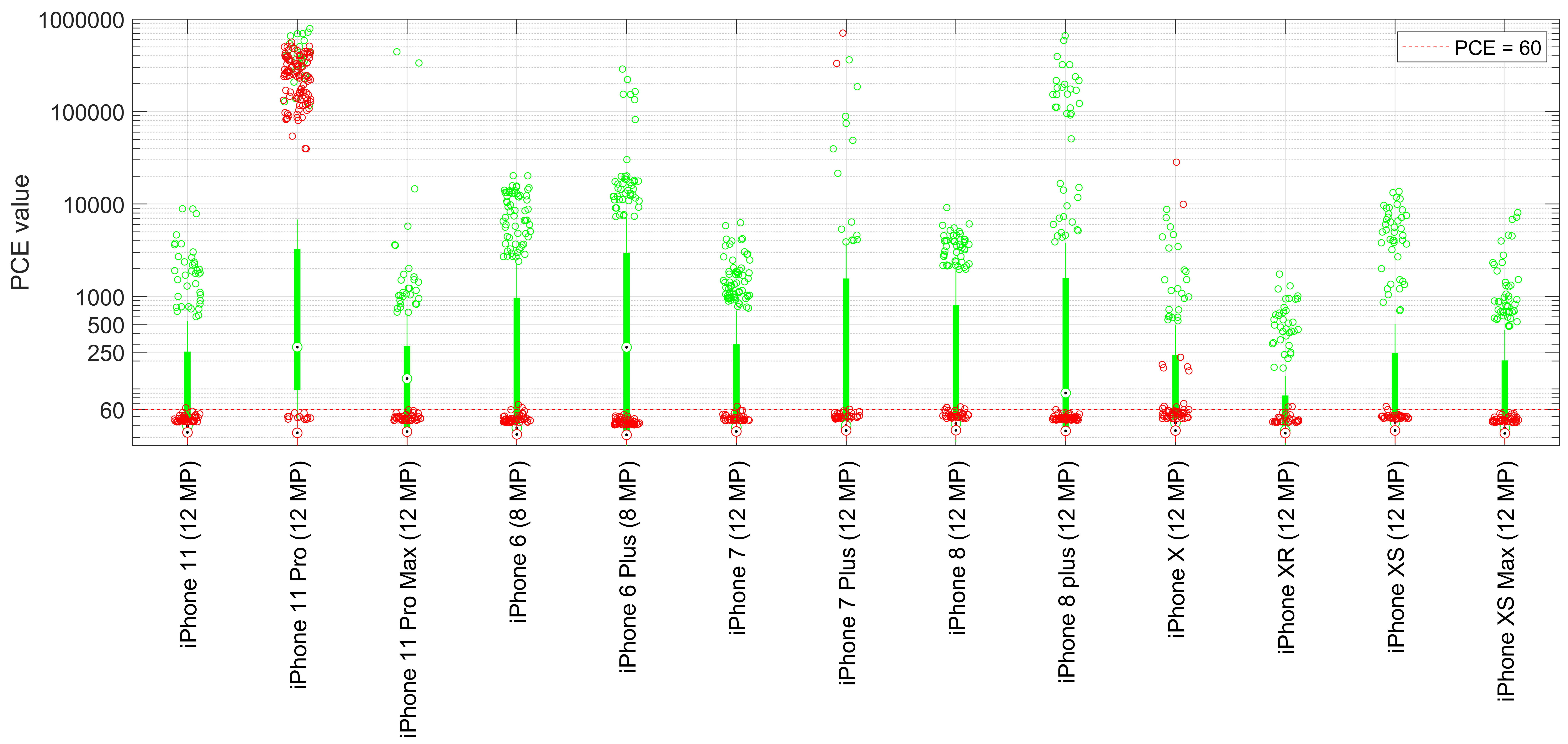}\\
    \includegraphics[width=.9\textwidth]{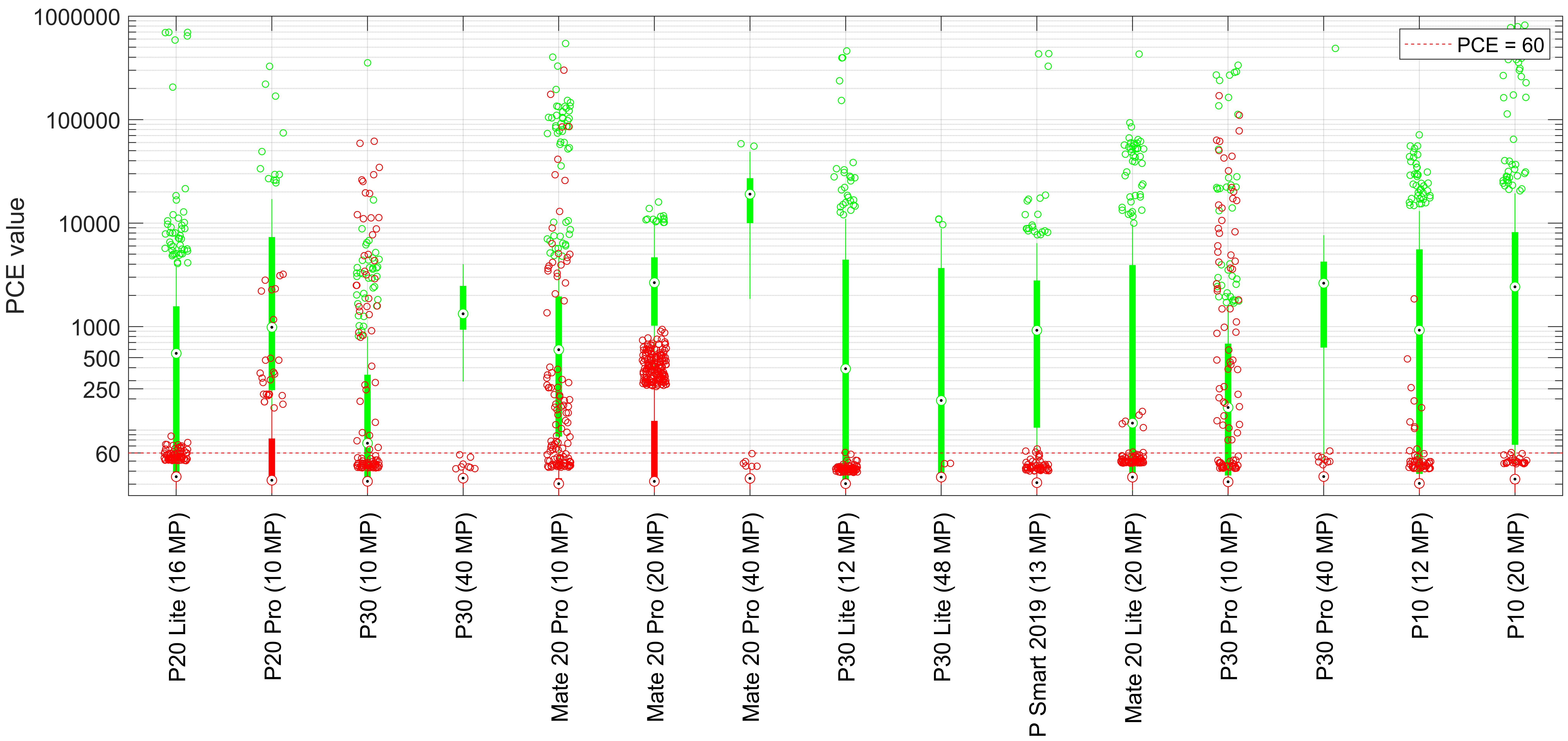}\\
    \includegraphics[width=.9\textwidth]{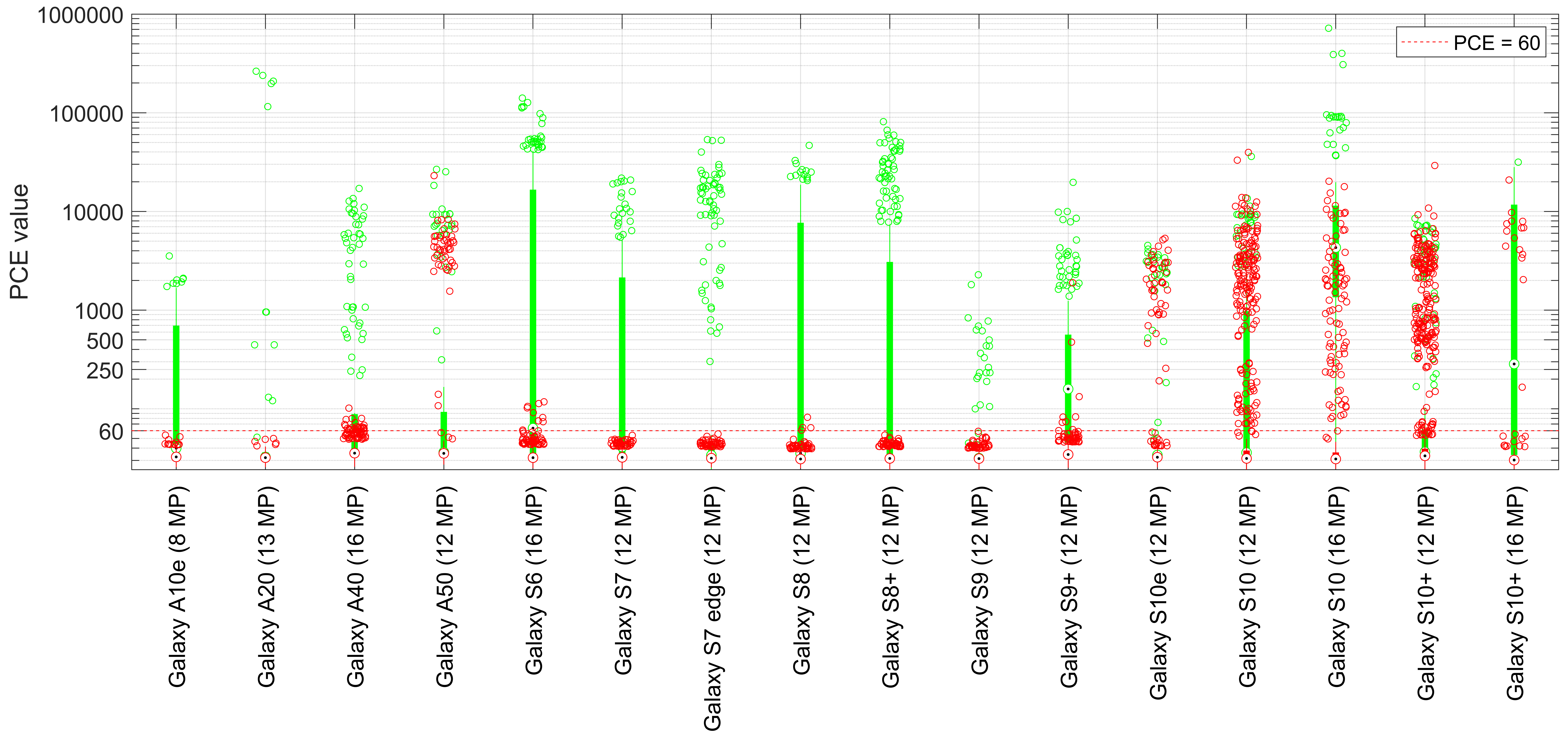}\\
  \end{tabular}
  \label{fig:apple_huawei_samsung}  
\end{figure*}

\begin{figure*}
\centering
\caption{PCE statistics computed from other smartphones (first plot) and from cameras (second plot). 
Matching and mismatching tests are reported in green and red, respectively. The threshold of $60$ is highlighted by the red dashed line.}
  \begin{tabular}{cc}
    \includegraphics[width=.9\textwidth]{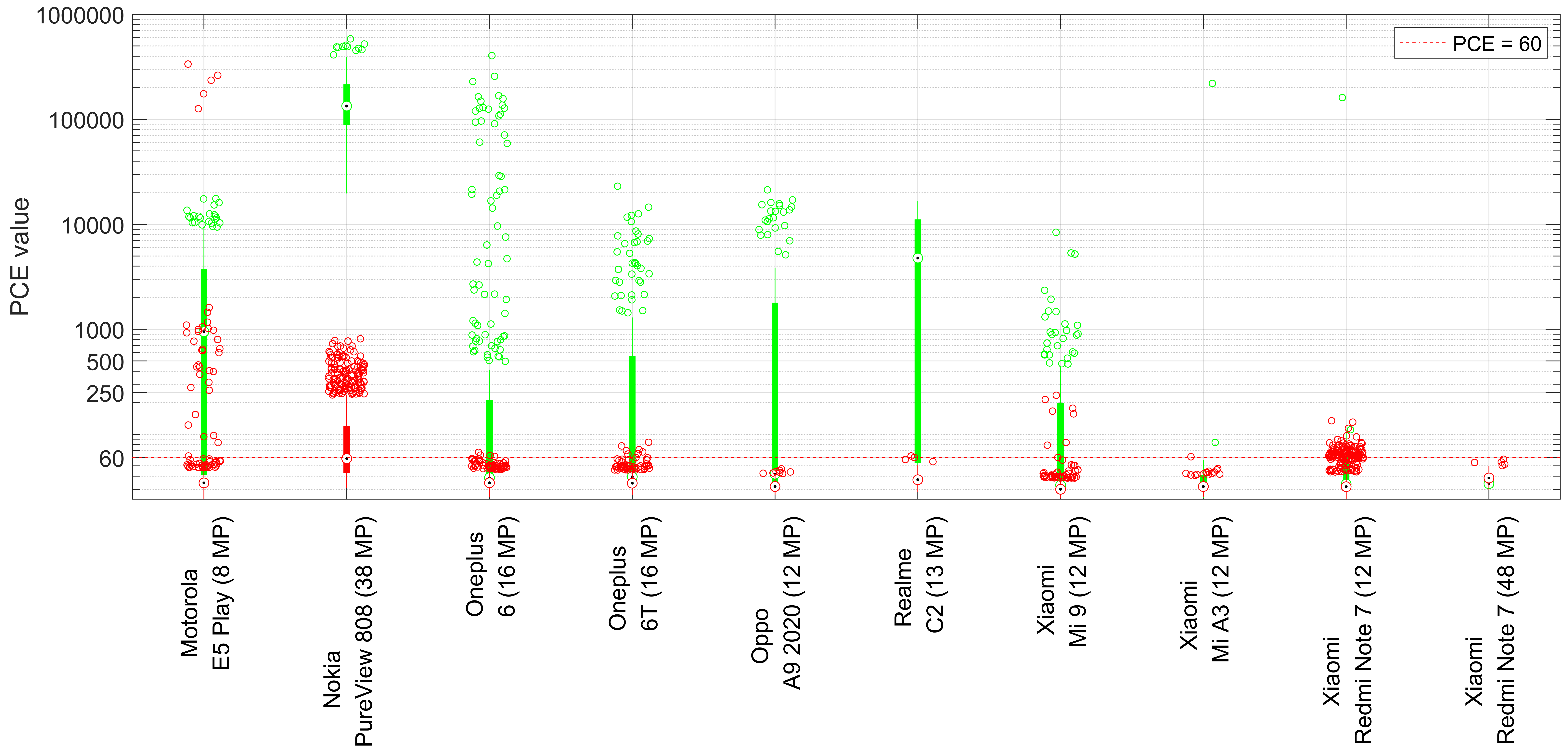}\\
\includegraphics[width=.9\textwidth]{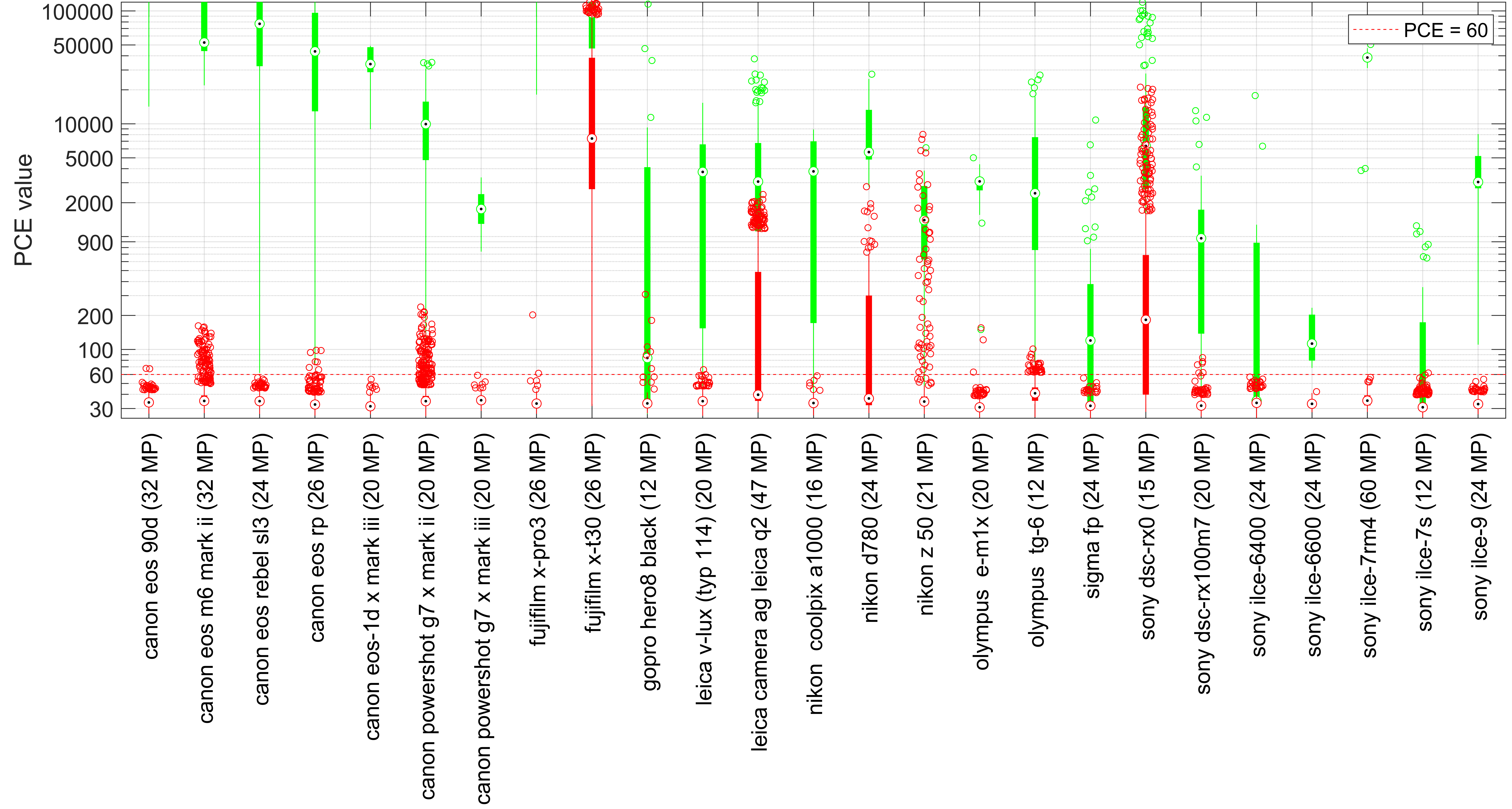}\\
  \end{tabular}
\label{fig:others_and_cameras}  
\end{figure*}

In Figures \ref{fig:apple_huawei_samsung} and \ref{fig:others_and_cameras}, we report the achieved PCE statistics: the first three plots cover Apple, Huawei, and Samsung devices, respectively.
The fourth plot includes smartphones from other brands, and the fifth plot is dedicated to compact and DSLR cameras.
Matching and mismatching tests are reported in green and red, respectively, while the red dashed line denotes a threshold value of $60$.
In Table \ref{tab:fpr_smartphones} and \ref{tab:rate_cameras} we also list the observed \fpr for smartphones and cameras, respectively.

At first glance, we notice that most brands seem to expose unexpected correlation patterns.
However, for Apple devices, the issue looks to be concentrated on a specific device, namely iPhone 11 Pro.
Conversely, several other brands' models expose significant \fpr values: Xiaomi Redmi Note 7 exceeds $5\%$; Galaxy A50 and Galaxy S10 exceed  $10\%$; P20 Pro and Mate 20 Pro exceed  $20\%$; Nokia Pure View 808, exceeds  $40\%$.
Cameras are also involved since Canon, Fujifilm, Leica, Sigma, and Nikon expose the same issue.
Some of them, like Fujifilm X-T30 and Sigma fp, also exceed $60\%$ of wrong attribution rate. Noticeably, the Fujifilm X-T30 embeds the camera serial number in the Exif metadata: it was thus possible to ensure that images from different Flickr users come from cameras with different serial numbers.
Furthermore, as the boxplots show, the PCE value obtained by non-matching images is often well above the median PCE obtained by matching images, suggesting that adjusting the threshold is not a solution to the problem.
These facts also suggest that even obtaining images from many exemplars of the same camera model could not sufficiently compute a reliable threshold with the classical Neyman-Pearson approach.


\begin{table}
\caption{Observed \fpr for smartphones. For some models, we detail the results achieved at each different image resolution.}
\centering
\begin{tabular}{lll}
\textbf{Apple devices}      & \fpr (\%)   \\ 
\hline
iPhone 11 (12 MP)           & 0.1            \\
iPhone 11 Pro (12 MP)       & 8.5            \\
iPhone 11 Pro Max (12 MP)   & 0              \\
iPhone 6 (8 MP)             & 0.1            \\
iPhone 6 Plus (8 MP)        & 0              \\
iPhone 7 (12 MP)            & 0.1            \\
iPhone 7 Plus (12 MP)       & 0.2            \\
iPhone 8 (12 MP)            & 0.2            \\
iPhone 8 plus (12 MP)       & 0              \\
iPhone X (12 MP)            & 0.8            \\
iPhone XR (12 MP)           & 0.2            \\
iPhone XS (12 MP)           & 0.1            \\
iPhone XS Max (12 MP)       & 0              \\
                            &                \\
\textbf{Huawei devices}     & \fpr (\%)   \\ 
\hline
P20 Lite (16 MP)            & 1.1            \\
P20 Pro (10 MP)             & 28.6           \\
P30 (10 MP)                 & 2.5            \\
P30 (40 MP)                 & 0              \\
Mate 20 Pro (10 MP)         & 4.2            \\
Mate 20 Pro (20 MP)         & 27.4           \\
Mate 20 Pro (40 MP)         & 0              \\
P30 Lite (12 MP)            & 0.1            \\
P30 Lite (48 MP)            & 0              \\
P Smart 2019 (13 MP)        & 0.2            \\
Mate 20 Lite (20 MP)        & 0.4            \\
P30 Pro (10 MP)             & 3.4            \\
P30 Pro (40 MP)             & 0.3            \\
P10 (12 MP)                 & 0.7            \\
P10 (20 MP)                 & 0.1            \\
                            &                \\
\textbf{Samsung devices}    & \fpr (\%)   \\ 
\hline
Galaxy A10e (8 MP)          & 0              \\
Galaxy A20 (13 MP)          & 0              \\
Galaxy A40 (16 MP)          & 3.2            \\
Galaxy A50 (12 MP)          & 10.8           \\
Galaxy S6 (16 MP)           & 0.7            \\
Galaxy S7 (12 MP)           & 0              \\
Galaxy S7 edge (12 MP)      & 0              \\
Galaxy S8 (12 MP)           & 0.3            \\
Galaxy S8+ (12 MP)          & 0              \\
Galaxy S9 (12 MP)           & 0              \\
Galaxy S9+ (12 MP)          & 0.6            \\
Galaxy S10e (12 MP)         & 5.1            \\
Galaxy S10 (12 MP)          & 18.1           \\
Galaxy S10 (16 MP)          & 13.5           \\
Galaxy S10+ (12 MP)         & 15.2           \\
Galaxy S10+ (16 MP)         & 4.6            \\
                            &                \\
\textbf{Other devices}      & \fpr (\%)   \\ 
\hline
Motorola E5 Play (8 MP)     & 2.5            \\
Nokia PureView 808 (38 MP)  & 48.4           \\
Oneplus 6 (16 MP)           & 0.2            \\
Oneplus 6T (16 MP)          & 0.6            \\
Oppo A9 2020 (12 MP)        & 0              \\
Realme C2 (13 MP)           & 1.5            \\
Xiaomi Mi 9 (12 MP)         & 0.7            \\
Xiaomi Mi A3 (12 MP)        & 0.2            \\
Xiaomi Redmi Note 7 (12 MP) & 7.1            \\
Xiaomi Redmi Note 7 (48 MP) & 0       
\end{tabular}
\label{tab:fpr_smartphones}  
\end{table}

\begin{table}
\centering
\caption{Observed \fpr for cameras.}
\label{tab:rate_cameras}
\begin{tabular}{ll}
\textbf{Camera devices}               & \fpr (\%)   \\ 
\hline
Canon EOS 90D (32 MP)                 & 0.2            \\
Canon EOS M6 Mark II (32 MP)          & 8.8            \\
Canon EOS Rebel SL3 (24 MP)           & 0              \\
Canon EOS RP (26 MP)                  & 0.6            \\
Canon EOS-1D X Mark III (20 MP)       & 0              \\
Canon PowerShot G7 X Mark II (20 MP)  & 6.2            \\
Canon PowerShot G7 X Mark III (20 MP) & 0              \\
Fujifilm X-Pro3 (26 MP)               & 1.4            \\
Fujifilm X-T30 (26 MP)                & 99.2           \\
Gopro Hero8 black (12 MP)             & 4.3            \\
Leica V-Lux (typ 114) (20 MP)         & 0.1            \\
Leica Q2  (47 MP)                     & 38             \\
Nikon Coolpix A1000  (16 MP)          & 0              \\
Nikon  D780 (13 MP)           & 38.5           \\
Nikon Z50  (21 MP)                    & 13.1           \\
Olympus E-M1X (20 MP)          & 0.3            \\
Olympus TG-6  (20 MP)                 & 2.8            \\
Sigma fp (24 MP)               & 0              \\
Sony DSC-Rx0 (24 MP)                      & 66.4           \\
Sony DSC-Rx100m7 (15 MP)                  & 0.8            \\
Sony ILCE-6400  (24 MP)               & 0              \\
Sony ILCE-6600 (24 MP)                & 0              \\
Sony ILCE-7rm4 (60 MP)                & 0              \\
Sony ILCE-7s (12 MP)                  & 0.1            \\
Sony ILCE-9 (24 MP)                   & 0             
\end{tabular}
\end{table}

\begin{table}[]
\caption{Summary of the main dataset's statistics}
\centering 
\begin{tabular}{  l r r }
	& \textbf{smartphones} & \textbf{cameras} \\ \hline
	\# collected images & 24.908 & 8.347\\
	\# devices          & 372 & 114 \\
	\# different models & 45 & 25 \\
	\# reference images &  12.752 & 3.586\\
	\# matching tests & 12.141 & 1.563 \\
	\# mismatching tests & 61.655 & 19.382 \\
\end{tabular}
\label{tab:summary}
\end{table}

%
\section{Inter Users Analysis}
\label{sec:discussion}

\begin{figure}[t]
\caption{PCE statistics among iPhone 11 Pro users. On the x-axis, each user's unique identifier is reported.
The fingerprint from user $101543825@N07$ is tested across all users.}
    \includegraphics[width=.48\textwidth]{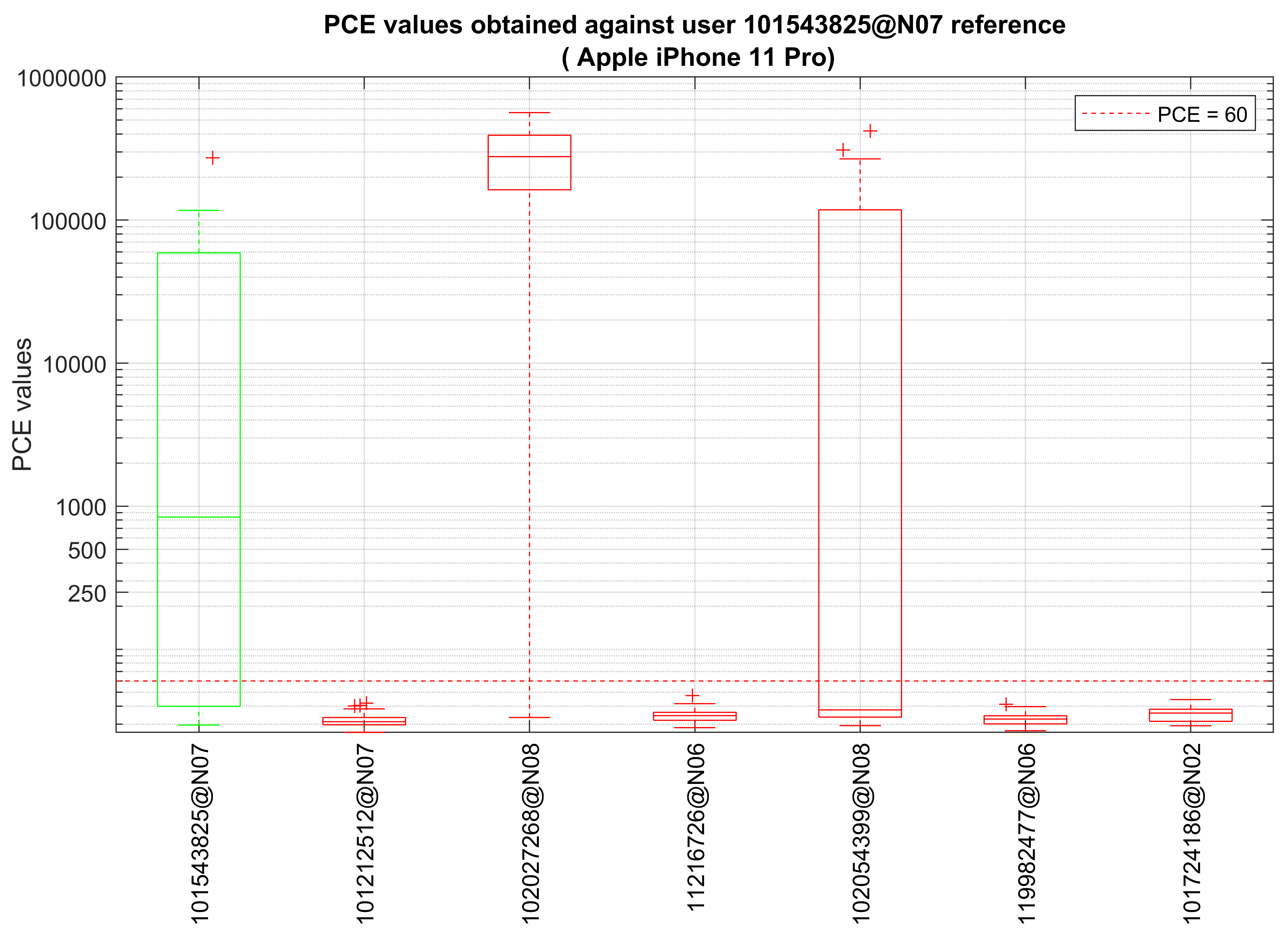}\\
\label{fig:breakdownApple}
\end{figure}

\begin{figure}[t]
\caption{Geolocalization tags of the Apple Phone 11 Pro colliding users.}
    \includegraphics[width=.45\textwidth]{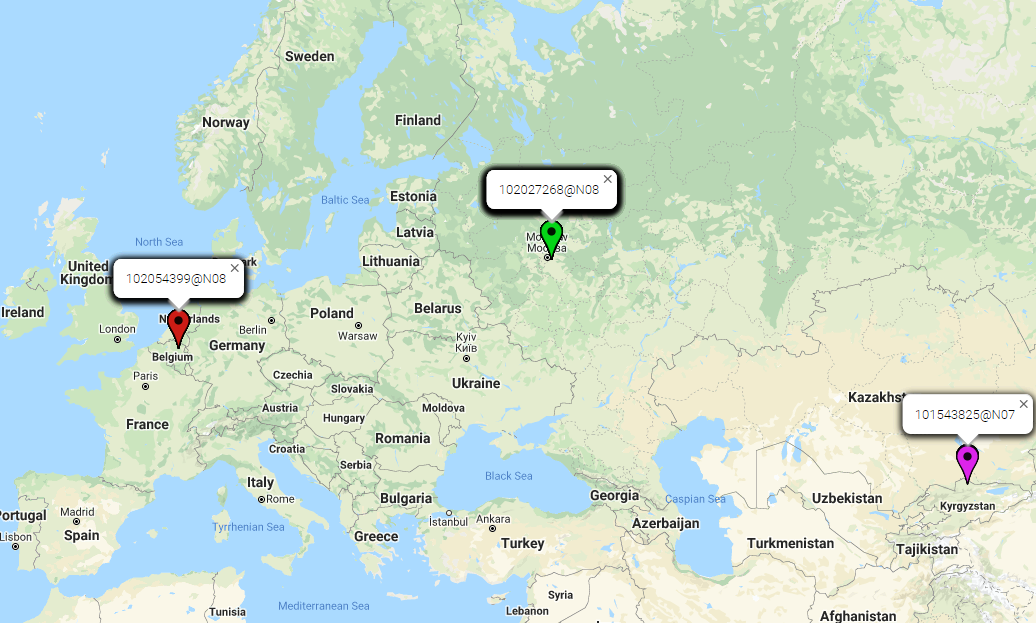}\\
\label{fig:geotagsApple}
\end{figure}

We analyzed, for each device producing unexpected correlation patterns, the achieved statistics within different users.
Concerning the Apple devices, in Figure \ref{fig:breakdownApple} we consider the fingerprint of user $101543825@N07$, and we show how the PCE ratios of the tested images are distributed across the users.
It can be clearly noticed that the unexpected matching affects three devices only.
The hypothesis that these users share the same phone, or that they can be attributed to the same person is considered strongly unlikely since image geolocalization tags expose distant acquisition places (see Figure \ref{fig:geotagsApple}).

A deeper analysis of the $Exif$ metadata highlights that most of the false alarms expose the tag \textit{Custom Rendered: Portrait HDR} highlighting that the images were captured in \textit{Portrait Mode}.
Similarly, a few reference images also expose the same tag.
It is known that such an acquisition mode introduces \nuas ~due to the in-camera background processing \cite{BaracchiIWDW20,albisani2021}.
The above results seem to be a reasonable explanation for the wrong attribution on Apple devices.

\begin{figure}[t]
\caption{PCE statistics among Samsung S10e users. On the x-axis, each user's unique identifier is reported.
The fingerprint from user $13346175@N08$ is tested across all users.}
    \includegraphics[width=.45\textwidth]{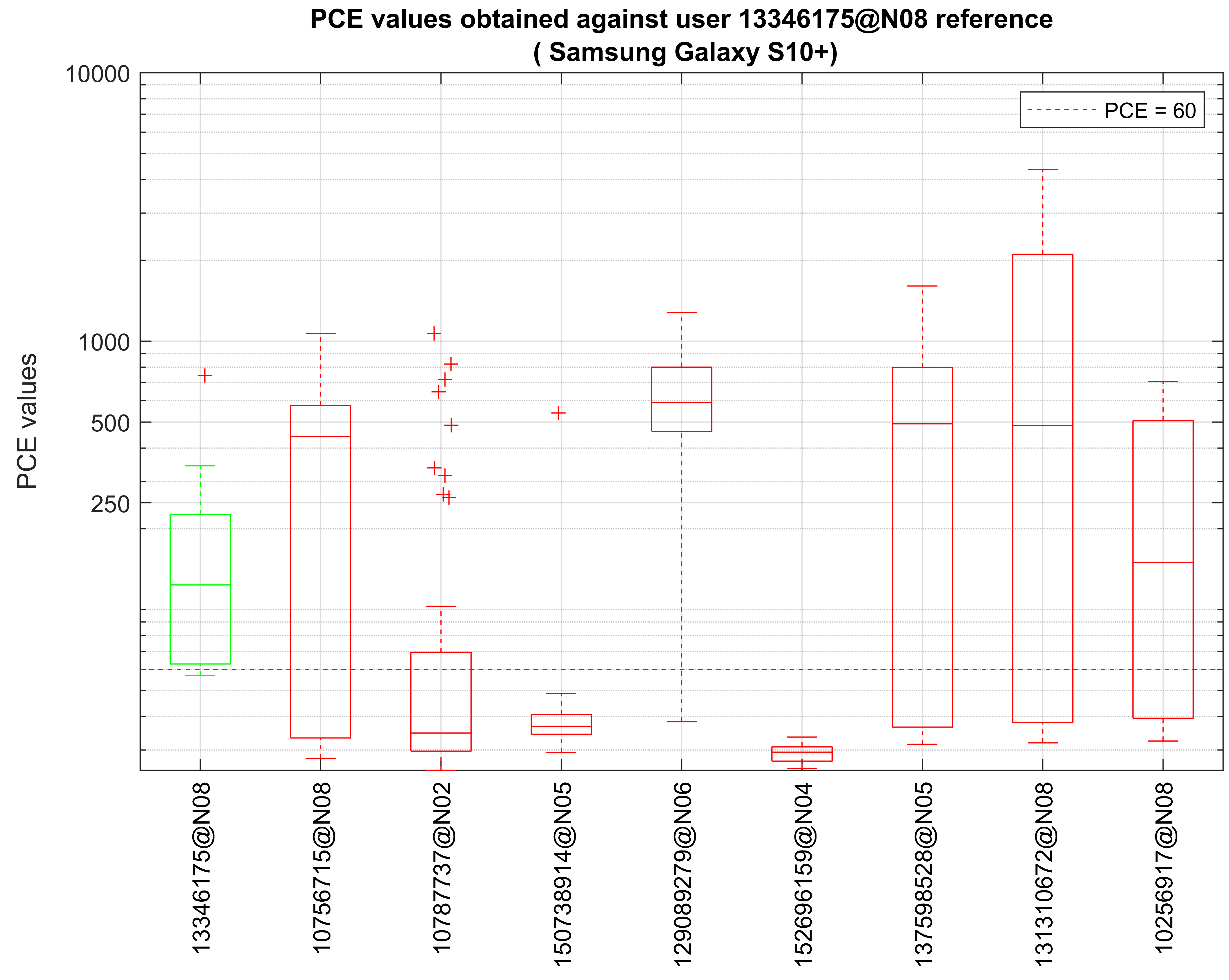}\\
\label{fig:breakdownSamsung}
\end{figure}

\begin{figure}[t]
\caption{Geolocalization tags of the Samsung S10e colliding users.}
    \includegraphics[width=.45\textwidth]{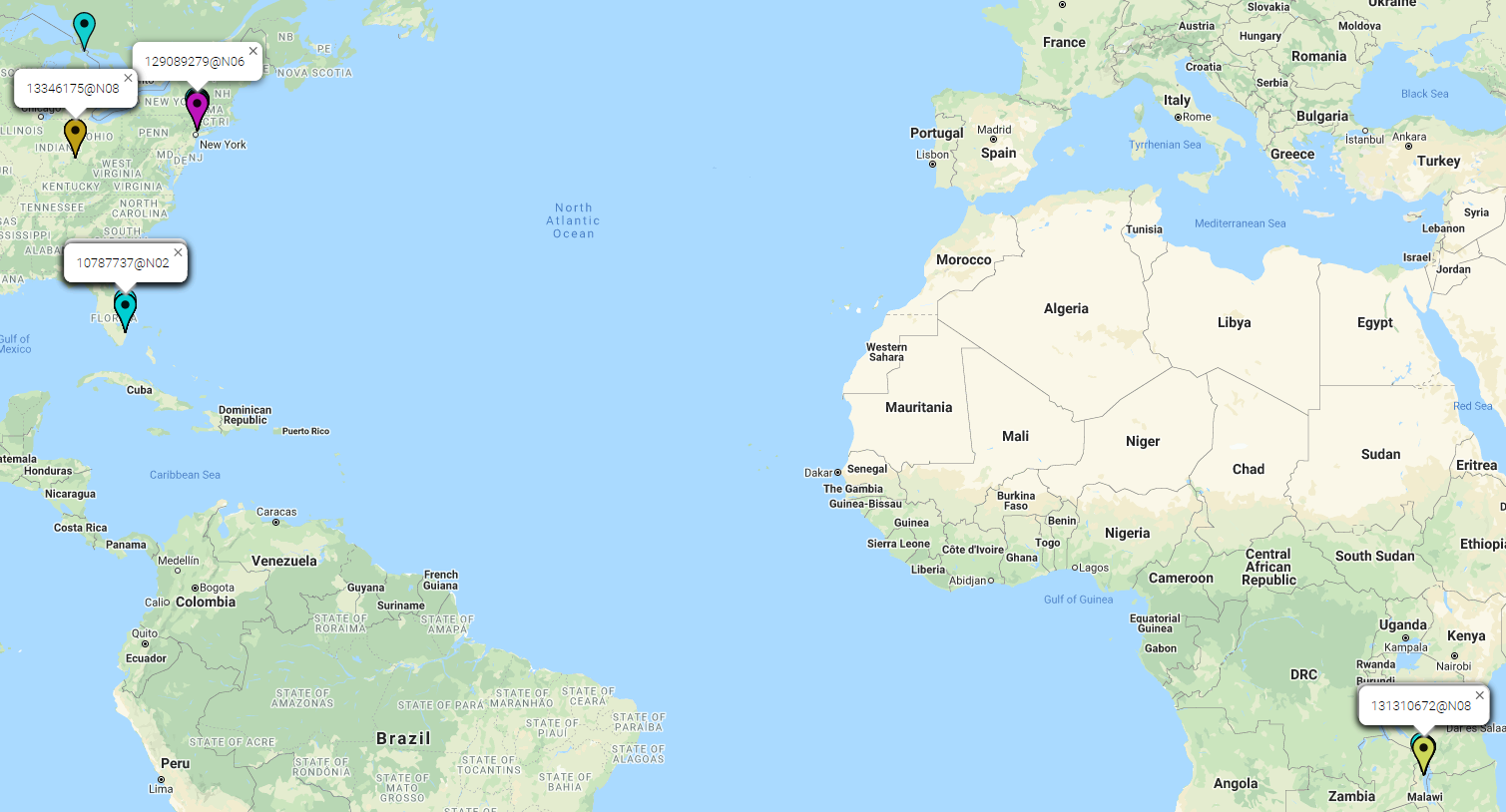}\\
\label{fig:geotagsSamsung}
\end{figure}

Similarly, in Figure \ref{fig:breakdownSamsung}, we report the Samsung S10e collisions for user $13346175@N08$'s fingerprint against all other users with the same device model.
It can be noticed that, in this case, false positives are spread across most users. Again, the geolocalization tags highlight that these images are reasonably acquired with different devices (Figure \ref{fig:geotagsSamsung}).
However, in this case, both image content and metadata analysis did not expose any evident anomaly (e.g., customized acquisition mode, shared firmware).
For the sake of brevity, we do not report similar plots for all the users; however, the cause of the wrong attribution on non-Apple devices does not seem to be explainable with some known processing.

It is worth mentioning that false positives are not independent of the resolution settings: the Huawei Mate 20 Pro, for example, suffers from false positives at the $10$MP and $20$MP resolution, but not at the maximum resolution.
This may suggest that the in-camera subsampling introduced by pixel binning could be the leading cause of the NUA.
This could be the case of Huawei P30 and Mate 20 Pro where false alarm only occurs on sub-sampled images.
However, Huawei P30 lite also adopts pixel binning but, in this case, \fpr is negligible independently on the resolution settings.

All the above facts strongly suggest that these unexpected correlation patterns cannot be attributed to specific processing that involves shooting modes, brand-proprietary technologies, and specific model in-camera processing.

\section{Conclusions}
\label{sec:conclusions}
In this paper, we investigated the reliability of PRNU-based source identification on modern devices.

We conducted a validation on a dataset obtained from Flickr, comprising $45$ smartphone and $25$ camera models. Results show that fingerprint uniqueness is not guaranteed for most brands.
Many models from popular brands (e.g., Huawei, Samsung) expose a false positive rate larger than 5\% when the commonly accepted threshold of 60 is employed.
A more in-depth analysis of image content and metadata highlighted that the cause of unexpected correlation patterns could not be reasonably attributed to a single specific imaging technology or processing. In contrast, it involves unusual shooting modes, brand-proprietary technology, and specific model in-camera processing. The achieved results also highlight that adjusting the threshold is not a solution to the problem.
We leave it for future work to make a deeper analysis of the strength, position, and a number of the correlation peaks for each device model to identify the possible leading causes and how we can possibly deal with them. Furthermore, we will consider extending the analysis of the problem to the same images exchanged through social media platforms.

Considering the widespread, worldwide application of this technology by law enforcement agencies, we believe it is of paramount importance to shed light on the issues raised in this paper.
Therefore, this work is to be intended as a call to action for the scientific community, which we invite to reproduce and validate our results and answer the questions that remained open.

\section*{Acknowledgment}
This work was partially supported by the Air Force Research Laboratory and by the Defense Advanced Research Projects Agency
under Grant HR00112090136, and by the Italian Ministry of Education, Universities and Research MIUR under Grant 2017Z595XS.

\bibliographystyle{IEEEbib}
\bibliography{strings}

\begin{IEEEbiography}[{\includegraphics[width=1in,height=1.25in,clip,keepaspectratio]{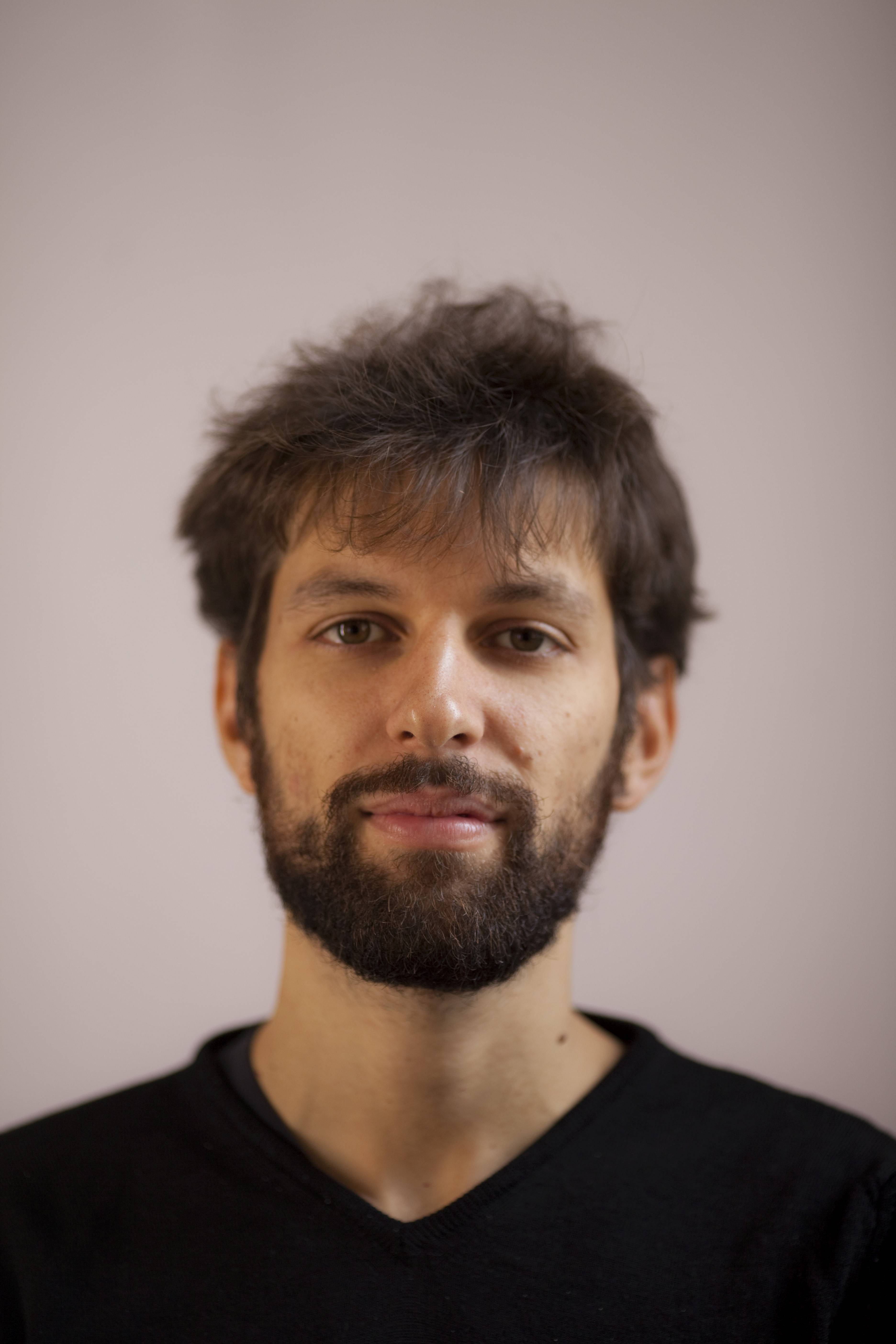}}]{Massimo Iuliani} graduated in Mathematics (summa cum laude) in 2011 at the University of Florence (Italy) and earned his Ph.D. in Mathematics in 2017 at the University of Florence under the supervision of Prof. Alessandro Piva, with a dissertation entitled “Image Forensics in the Wild”. He works as an assistant researcher in the field of image processing for security and forensic applications at the Department of Information Engineering of the University of Florence, Italy. He also works as technical supervisor at FORLAB, the Multimedia Forensics Laboratory (www.forlab.org) at PIN s.c.r.l. Educational and Scientific Services for the University of Florence. His main activities involve the training of law enforcement and legal operators and the consultancy in the multimedia forensics field.
\end{IEEEbiography}

\begin{IEEEbiography}[{\includegraphics[width=1in,height=1.25in,clip,keepaspectratio]{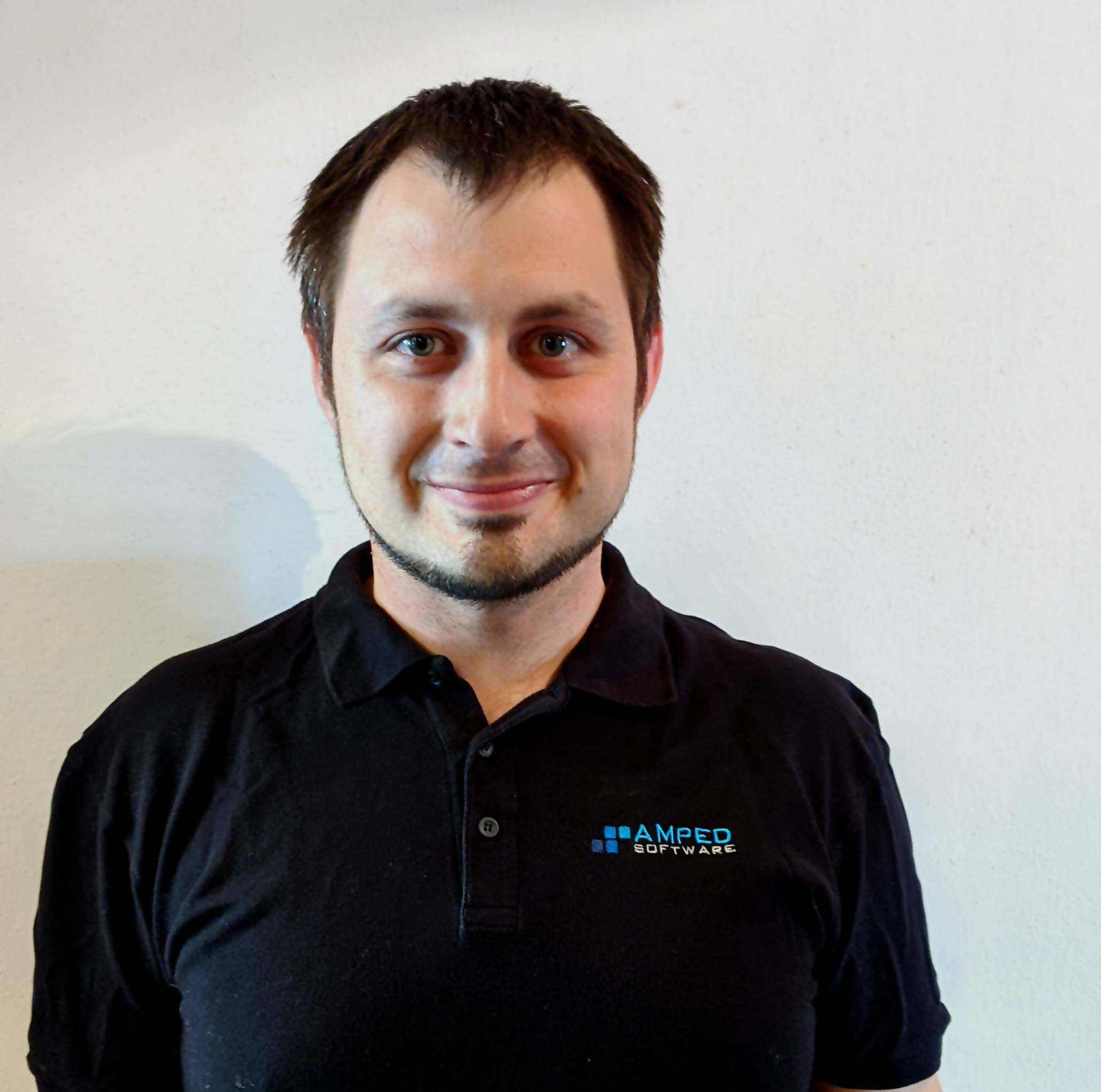}}]{Marco Fontani} graduated in Computer Engineering in 2010 at the University of Florence and earned his Ph.D. in Information Engineering in 2014 at the University of Siena. He is currently Product Manager at Amped Software, and he is member of the IEEE Information Forensics and Security Technical Committee. He participated in several research projects, funded by the EU and by the EOARD. He is the author or co-author of several journal papers and conference proceedings. He has also experience in delivering training to law enforcement and he provided expert witness testimony on several forensic cases involving digital images and videos. 
\end{IEEEbiography}

\begin{IEEEbiography}[{\includegraphics[width=1in,height=1.25in,clip,keepaspectratio]{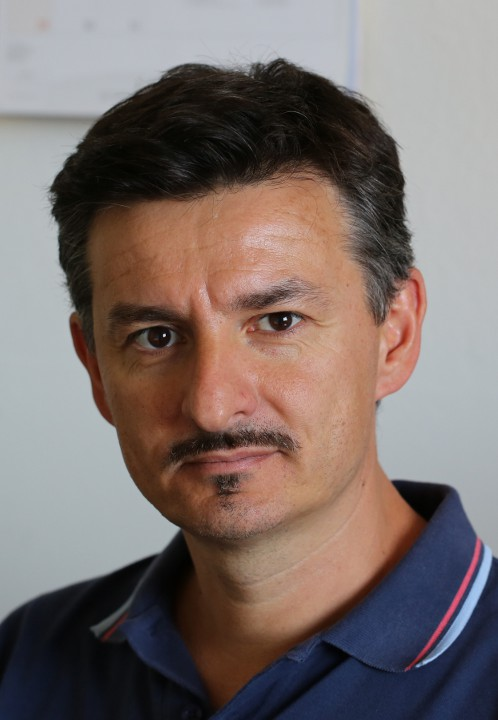}}]{Alessandro Piva} is Associate Professor at the Department of Information Engineering of the University of Florence. He is also head of FORLAB – Multimedia Forensics Laboratory of the University of Florence. His research interests lie in the areas of Information Forensics and Security, and of Image and Video Processing. In the first topic, he was interested in data hiding, signal processing in the encrypted domain, image and video forensic techniques. In the second area, he was interested in the design of image and video processing and analysis techniques for Cultural Heritage, medical and industrial applications. In the above research topics he has been co-author of more than 50 papers published in international journals and 120 papers published in international conference proceedings. He holds 3 Italian patents and an International one on digital watermarking. He is IEEE Fellow, and he is member of the IEEE Information Forensics and Security Technical Committee.
\end{IEEEbiography}

\EOD{}

\end{document}